\documentclass[epsfig,psfig,aps,twocolumn,prb,showpacs,notitlepage,tightenlines,10pt]{revtex4-2}
\usepackage{graphicx, color}
\usepackage{pbox}
\usepackage{verbatim}
\usepackage{mwe}
\usepackage{amsmath, amsthm, amssymb}
\usepackage{float}
\usepackage{framed}
\usepackage{empheq}
\usepackage{MnSymbol}
\usepackage{bm}
\usepackage{microtype}
\usepackage{bbold}
\usepackage[colorlinks=true,citecolor=red,urlcolor=blue]{hyperref}
\begin{document}

\title{Quantum geometric scattering of a Dirac particle by a Berry curvature domain wall}

\author{Lassaad Mandhour $^{1}$}
\email{lassaad.mandhour@istmt.utm.tn}
\author{Fr\'ed\'eric Pi\'{e}chon $^{2}$}
\email{frederic.piechon@universite-paris-saclay.fr}
\affiliation{
$^1$Laboratoire de Physique de la Mati\`ere Condens\'ee, Facult\'e des Sciences de Tunis, Universit\'e Tunis El Manar, Campus Universitaire 1060 Tunis, Tunisia\\
$^2$  Universit\'e Paris-Saclay, CNRS, Laboratoire de Physique des Solides, 91405 Orsay Cedex, France}
\date{\today}
\begin{abstract}
We investigate the scattering of a three-dimensional massless Dirac particle through a domain wall separating two regions with identical energy spectra but distinct Berry curvature dipoles. We demonstrate that the quantum geometric mismatch induces partial reflection and transmission despite identical incident and refracted momenta. These results highlight the role of engineered quantum geometric interfaces as key tools to control Dirac particle scattering.

\end{abstract}
\maketitle

\section{Introduction}

In condensed matter physics, excitations in numerous materials exhibit behavior analogous to two-dimensional (2D) or three-dimensional (3D) massless Dirac quasiparticles \cite{Armitage2018,Lv2021,McClarty2022}.
Similarly, engineered band structures featuring 2D and 3D Dirac cone crossings have been demonstrated in a variety of photonic and phononic crystals, as well as in metamaterials \cite{Ozawa2019,Zhang2018,Yang2024}. Additionally, cold atomic gases in optical lattices \cite{Cooper2019} and exciton-polaritons in metamaterials provide versatile experimental platforms for exploring 2D Dirac physics \cite{Ozawa2019,Jaquemin2014}.
Massless Dirac particles exhibit a characteristic linear Dirac cone dispersion. Their pseudospinor wavefunctions satisfy an effective multiband eigenvalue equation $(c{\bm p} \cdot {\boldsymbol \Lambda})\psi=E({\bm p}) \psi$, where ${\boldsymbol \Lambda}$ are system-dependent  pseudospin matrices \cite{Dirac1928} which encode the quantum geometry of the wavefunctions, including the Berry connection, Berry phase, and Berry curvature. It is now well established that these quantum geometric propoerties play a crucial role in determining the physical properties of the Dirac quasiparticles \cite{Provost1980,Berry1984,Lin2021,Graf2021,Mera2022,Graf2023}.

As mentioned above, the quantum geometry, particularly the Berry curvature, in massless Dirac materials depends on the momentum-space pseudospin texture encoded in the wavefunctions and the structure of the effective Hamiltonian, including system-specific pseudospin matrices. 
In systems such as graphene and the $\alpha-T_3$ model, time-reversal and inversion symmetries enforce a vanishing Berry curvature throughout the Brillouin zone, except at the Dirac points where it becomes ill-defined.
In contrast, 3D massless topological chiral semimetals \cite{Lv2021,Armitage2018} are characterized by band-touching points that carry Berry-curvature monopoles \cite{Berry1984}. Another class of systems with finite Berry curvature is provided by massless multifold Hopf semimetals, which feature linear multifold band crossings giving rise to a dipolar Berry-curvature structure \cite{Graf2023}.

The influence of pseudospin and more generally the underlying quantum geometry on the transport properties of Dirac particles is often revealed through their scattering behavior at potential barriers. This pseudospin degree of freedom plays an important role in determining transmission and reflection probabilities, leading to distinctive phenomena such as Klein tunneling, where perfect transmission occurs at normal incidence \cite{Katsnelson2006,Allain2011}. Extensions of this framework have explored barriers that combine electric and magnetic potentials, further emphasizing the importance of the Berry phase \cite{Bouhadida2020}. Beyond these barriers, a variety of scattering mechanisms have been studied in massless Dirac systems, including lattice strain \cite{Yesilyurt, Hung, Zhai, Fujita, Zhai2010, Islam}, line defects \cite{Cheng, Gunlycke, Rodrigues2012, Rodrigues2013, Rodrigues2016, Paez2015}, velocity barrier \cite{Raoux2010, Concha2010}, boundaries between graphene regions exhibiting rotated crystallographic axes \cite{Romeo2018}, and twist-angle domain wall in twisted bilayer graphene \cite{Padhi2020}. In all cases, scattering of Dirac particles results from external perturbations that induce spatial variations in the energy spectrum and/or momentum, leading to effective pseudospin scattering. In contrast, in our recent work \cite{Mandhour2025}, we have investigated the scattering of Dirac particles through a domain wall that separates two regions with identical energy spectra but distinct Berry phases. It demonstrated that a quantum geometric mismatch at a domain wall leads to a partial reflection-transmission of the Dirac particle.

Building on this framework, the present study focuses on scattering at a domain wall that highlights the impact of a Berry curvature dipole mismatch. Specifically, we analyze a three-band model of massless multifold Hopf semimetals with a tunable Berry curvature dipole \cite{Graf2023}. The domain wall separates two regions that share identical energy spectra but differ in the orientation of their Berry dipole vector, allowing us to investigate how this geometric mismatch influences the scattering properties.

The paper is organized as follows. In section \ref{sectionII}, we introduce a three-band tight-binding model on the cubic lattice that features a tunable Berry curvature dipole. In section \ref{sectionIIIA}, we employ a continuum low-energy description to calculate the transmission probability across a Berry curvature domain wall. Section \ref{sectionIIIB} revisits the problem using the tight-binding framework on the cubic lattice. Finally, section \ref{sectionIV} summarizes and concludes the present work.

\section{Three dimensional massless Dirac particle with a tunable Berry curvature Dipole}
\label{sectionII}

\subsection{Three bands tight-binding model on the cubic lattice}

We consider a cubic lattice with three atomic orbitals per site. Using a second-quantized notation, 
we define the creation operators $X^{\dagger}_{{\bf R}}\equiv |X_{{\bf R}}\rangle$ for orbital $X={A,B,C}$ at position
${\bf R}=a(n_x  \bm{e}_x+n_y  \bm{e}_y +n_z  \bm{e}_z)$, where $a=1$ denotes the interatomic distance.
The homogeneous tight-binding Hamiltonian is expressed as
\begin{equation}
 \begin{array}{ll}
H_\alpha=   t/2 \sum_{{\bf R}}&[A^{\dagger}_{{\bf R}}(B_{{\bf R}+\bm{e}_x}+B_{{\bf R}-\bm{e}_x})\\
&-c_\alpha A^{\dagger}_{{\bf R}}(B_{{\bf R}+\bm{e}_z}-B_{{\bf R}-\bm{e}_z})\\
&+C^{\dagger}_{{\bf R}}(B_{{\bf R}+\bm{e}_y}+B_{{\bf R}-\bm{e}_y})\\
&+s_\alpha C^{\dagger}_{{\bf R}}(B_{{\bf R}+\bm{e}_z}-B_{{\bf R}-\bm{e}_z})+\textrm{H.c.}],
\end{array}
\end{equation}
with $s_\alpha=\sin \alpha$ and $c_\alpha=\cos \alpha$.
 The parameter $\alpha$ governs the relative strength of the hopping amplitudes between $B-A$ and $B-C$ sites along the $z$ direction. In the limiting cases $\alpha=0$ and $\alpha=\pi/2$, the hopping is restricted to the $B-A$ and $B-C$ bonds, respectively. Consequently, all physically inequivalent configurations are fully characterized by restricting $\alpha$ to the interval $0 \leq \alpha \leq \pi/2$.
 
\begin{figure}[]
       \centering
    \includegraphics[width=0.5\textwidth]{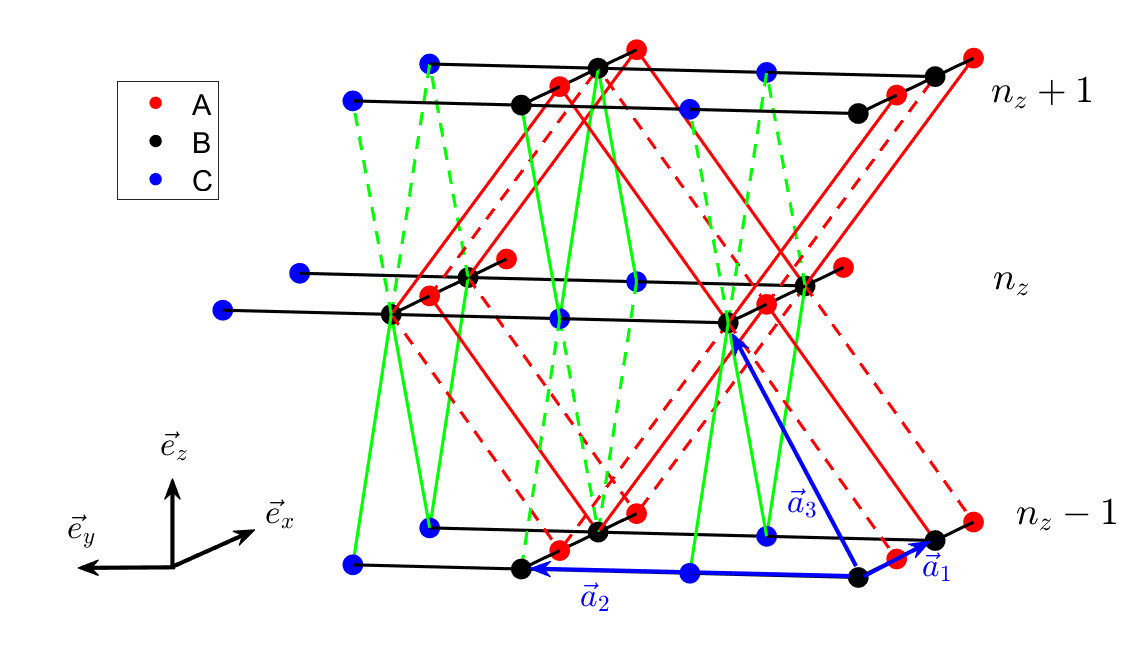}
    \includegraphics[width=0.23\textwidth]{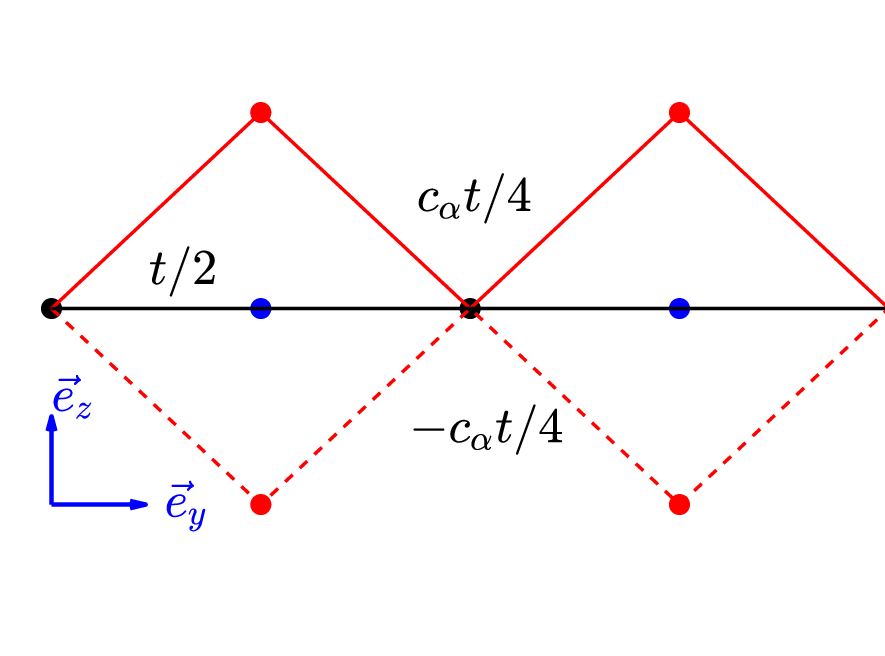}
    \includegraphics[width=0.24\textwidth]{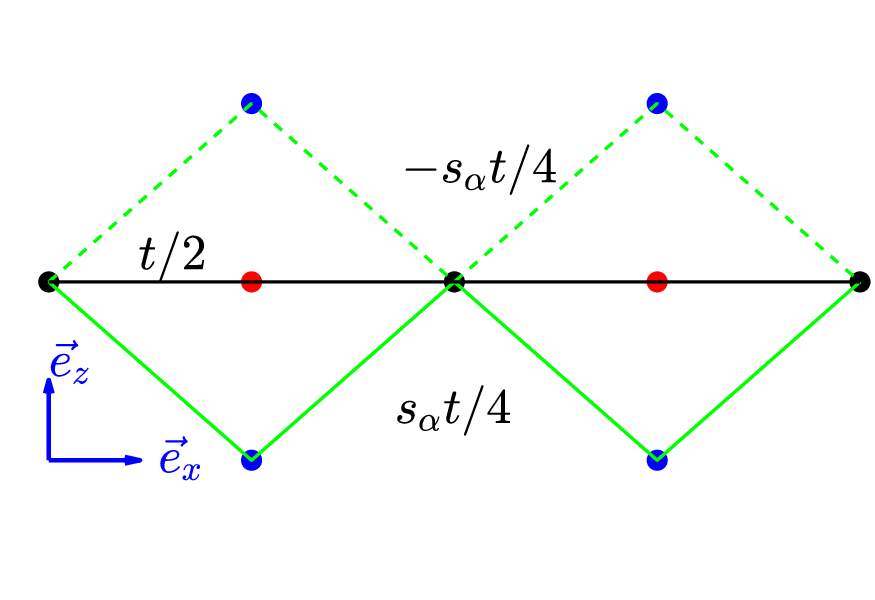}
 \caption{Top panel: Schematic representation of the $AB$ stacking of layers of 2D Lieb lattice. Bottom panel: Cut of the lattice at $x=a/2$ on the left and at $y=a/2$ on the right. The hopping amplitudes between $B$ and $A$ sites is $t/2$ along $x$ direction and alternating alternating hoppings $\pm c_\alpha t/4$ along $z$ direction. The hopping amplitudes between $B$ and $C$ sites is $t/2$ along $y$ direction and alternating hoppings $\pm s_\alpha t/4$ along $z$ direction. $\vec{a}_1=a(2,0,0)$, $\vec{a}_2=a(0,2,0)$ and $\vec{a}_3=a(1,1,1)$ are the Bravais lattice
vectors.}
     \label{figure0}
\end{figure}

The Hamiltonian has a chiral symmetry in the orbital basis and anticommutes with the diagonal matrix $S=\textrm{diag}(1,-1,1)$.
Because of the latter, the energy spectrum is  particle-hole symmetric and there is a flat band at zero energy.
Another way to obtain the same model consists to build a periodic $AB$ stacking of layers of 2D Lieb lattice as depicted in Fig. \ref{figure0}.
Experimentally, this model can be realized with ultracold atoms in an optical lattice \cite{Fulga2018}. The hopping amplitudes, and hence the parameter $\alpha$, can be tuned by adjusting the laser parameters. This controllability enables precise engineering of the lattice geometry and hopping strengths.

\begin{figure}[]
       \centering
    \includegraphics[width=0.45\textwidth]{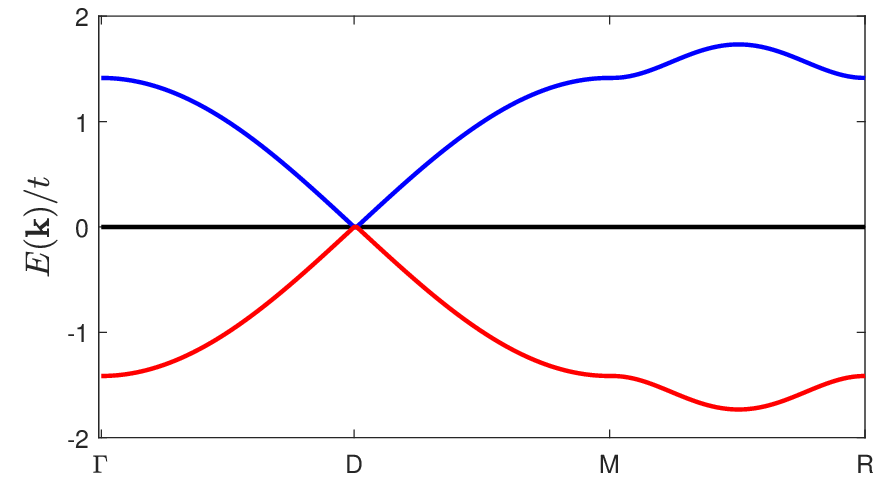}
 \caption{Energy band spectrum along the path $\Gamma(0,0,0)\rightarrow D(\pi/2,\pi/2,0)\rightarrow M(\pi,\pi,0)\rightarrow R(\pi,\pi,\pi)$.}
     \label{figure11}
\end{figure}

Defining the three components Bloch creation operators $d^{\dagger}_{\bm k}=\sum_{{\bf R}}e^{i {\bm k \cdot R}}d^{\dagger}_{\bm R}\equiv (A^{\dagger}_{{\bf k}},B^{\dagger}_{{\bf k}},C^{\dagger}_{{\bf k}})$
we can rewrite the Hamiltonian
$H_\alpha=\int_{\textrm{BZ}} \frac{d{\bm k}}{(2\pi)^3}d^{\dagger}_{\bm k}h_\alpha({\bm k})d_{\bm k}$
with the Bloch Hamiltonian matrix $h_\alpha({\bm k})$ given by
\begin{equation}
 h_\alpha({\bm k}) =t\left(
  \begin{array}{lll}
   0&c_x -i c_\alpha s_z&0\\
   c_x +i c_\alpha s_z&0& c_y -i s_\alpha s_z\\
   0&c_y +i s_\alpha s_z&0
  \end{array}
\right),
\end{equation}
with $c_{x,y}=\cos k_{x,y}$ and $s_{z}=\sin k_z$.
The energy band spectrum is given by $E_s({\bm k})=s t\sqrt{c_x^2 +c_y^2+s_z^2}$, where $s=-,0,+$ (see Fig. \ref{figure11}).
The dispersive bands $E_\pm({\bm k})$ touch the flat band $E_0=0$ at 3D massless Dirac points located at
\begin{equation}
{\bm D}_{\xi_x,\xi_y,\xi_z}=\left(\xi_x \pi/2,\xi_y \pi/2, (1-\xi_z) \pi/2\right),
\label{dirac_point}
\end{equation}
where $\xi_{x,y,z}=\pm$ denote the eight Dirac points in the first Brillouin zone.
Interestingly, despite its explicit presence in the Hamiltonian, the parameter $\alpha$ does not affect the energy spectrum.

The band eigenstates are given by $d_{s,{\bm k}}^{\dagger}=\sum_{X=A,B,C} \psi_s^{X}X^{\dagger}_{{\bm k}}$.
By introducing the three-component eigenstate $\psi_{s}({\bm k})\equiv (\psi_s^{A},\psi_s^{B},\psi_s^{C})$ we obtain
\begin{equation}
\begin{array}{l}
\psi_{\pm}({\bm k})=\frac{1}{\sqrt{2}\sqrt{ c_x^2 +c_y^2+s_z^2}}\left(
  \begin{array}{c}
   c_x - i c_\alpha s_z\\
  \pm \sqrt{c_x^2 +c_y^2+s_z^2}\\
   c_y +i s_\alpha s_z
  \end{array}\right),\\
  \psi_{0}({\bm k})=\frac{1}{\sqrt{ c_x^2 +c_y^2+s_z^2}}\left(
  \begin{array}{c}
   -(c_y -i s_\alpha s_z)\\
   0\\
  c_x  +i c_\alpha s_z
  \end{array}\right).
  \end{array}
\end{equation}
In contrast to the energy spectrum, the wavefunctions explicitly depend on the parameter $\alpha$.

The Berry curvature of each band is a $\alpha$ dependent pseudovector ${\bm \Omega}_{s}^\alpha({\bm k})$ given by
\begin{equation}
 {\bm \Omega}_{s}^\alpha({\bm k})=\kappa_s \frac{s_\alpha c_x+c_\alpha c_y}{(c_x^2+c_y^2+s_z^2)^2}\left(
  \begin{array}{l}
   -c_x s_y c_z\\
   -s_x c_y c_z\\
  s_x s_y s_z
  \end{array}\right).
  \label{berry_curv}
\end{equation}
where $\kappa_\pm=-1$ and $\kappa_0=2$, with $c_{x,y,z}=\cos k_{x,y,z}$ and $s_{x,y,z}=\sin k_{x,y,z}$. 

\subsection{Continuum low-energy model}
The continuum low-energy effective Hamiltonian model around the various Dirac points
${\bm k}={\bm D}_{\xi_x,\xi_y,\xi_z}+{\bm q}$ takes a Dirac like form with a tunable effective pseudospin:
\begin{equation}
\begin{array}{ll}
  H_\alpha({\bm q})&=\hbar v_F(\hat{q}_x \Lambda_x+ \hat{q}_y \Lambda_y+\hat{q}_z \Lambda_z^{\alpha})\\
  &=\hbar v_F \left(
  \begin{array}{ccc}
   0 &  \hat{q}_x-ic_\alpha \hat{q}_z &0\\
  \hat{q}_x+ic_\alpha \hat{q}_z &  0 & \hat{q}_y-is_\alpha \hat{q}_z   \\
   0 &  \hat{q}_y+is_\alpha \hat{q}_z &0
  \end{array}
  \right),
  \end{array}
 \label{Hamilton0}
 \end{equation}
where $v_F=at/\hbar$ is the Fermi velocity and
\begin{eqnarray}
\hat{q}_{x,y} = -\xi_{x,y} q_{x,y},\hspace{1cm} \hat{q}_{z}=\xi_{z} q_{z}.
\label{q_xyz}
\end{eqnarray}
The pseudospin matrices are given by
\begin{equation}
  \Lambda_x=\left(
  \begin{array}{ccc}
   0 &  1 &0\\
  1 &  0 & 0   \\
   0 & 0 &0
  \end{array}
  \right),
  \Lambda_y=\left(
  \begin{array}{ccc}
   0 &  0&0\\
  0 &  0 & 1   \\
   0 & 1 &0
  \end{array}
  \right),
  \Lambda_z^\alpha=i\left(
  \begin{array}{ccc}
   0 &  -c_\alpha &0\\
  c_\alpha &  0 & -s_\alpha   \\
   0 & s_\alpha &0
  \end{array}
  \right),
\end{equation}

\begin{figure}[h]
       \centering
    \includegraphics[width=0.5\textwidth]{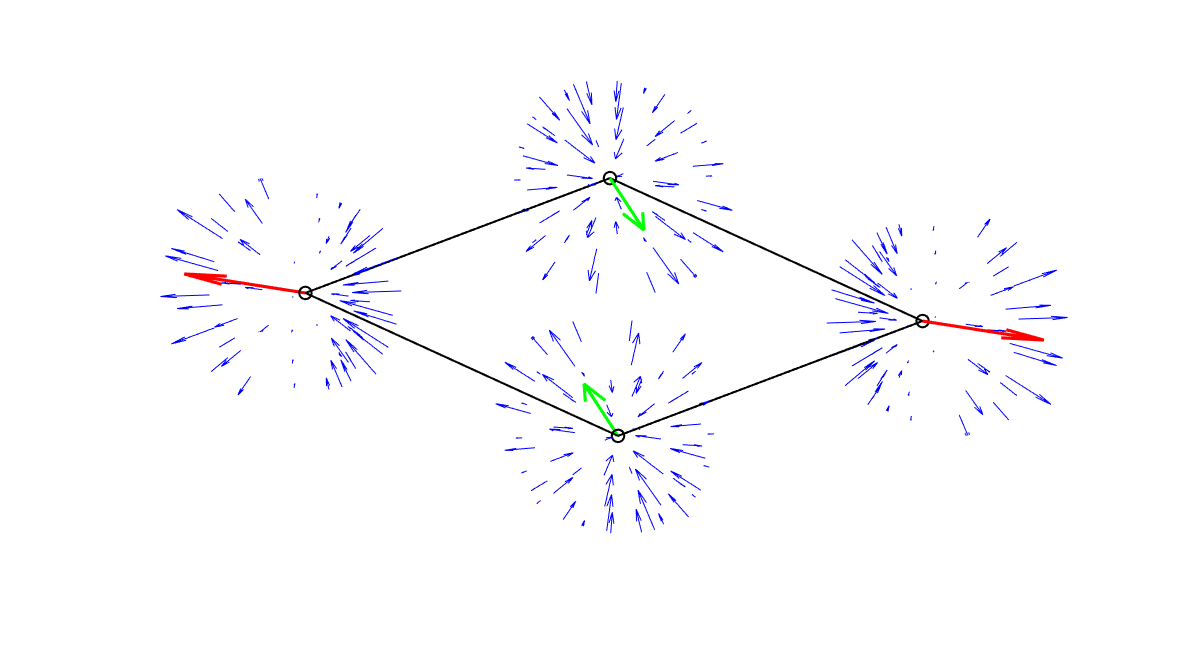}
 \caption{Vector field plots of the Berry curvature (blue arrows) and Berry dipole (Green arrows for $\xi_x\xi_y=1$, red arrows for $\xi_x\xi_y=-1$) of the conduction band near the four Dirac points ${\bm D}_{\xi_x=\pm 1,\xi_y=\pm 1,\xi_z=1}$.}
     \label{figure_dip}
\end{figure}

The energy spectrum is composed of two dispersive bands $E_s(q)=s\hbar v_F\left | {\bm q} \right |$ ($s=\pm$) that form a 3D Dirac cone, and a flat band with energy $E_0=0$.
The wave functions depend explicitly on the tuning parameter $\alpha$ and write as:
\begin{equation}
\psi_{s}({\bm k})=\frac{1}{\sqrt{2}|{\bm q}|}\left(
  \begin{array}{c}
   \hat{q}_x - i c_\alpha \hat{q}_z\\
   s |{\bm q}|\\
   \hat{q}_y+i s_\alpha \hat{q}_z
  \end{array}\right),
  \psi_{0}({\bm k})=\frac{1}{ | {\bm q}| }\left(
  \begin{array}{c}
   -(\hat{q}_y-i s_\alpha \hat{q}_z)\\
   0\\
  \hat{q}_x +i c_\alpha \hat{q}_z
  \end{array}\right).
\end{equation}
In the vicinity of the Dirac point ${\bm D}_{\xi_x,\xi_y,\xi_z}$, the Berry curvature, given by Eq. (\ref{berry_curv}), takes the form of a Berry dipole, as first noted in \cite{Graf2023} 
\begin{equation}
{\bm \Omega}_{s}^\alpha({\bm q})=\kappa_s \frac{({\bm d}_\alpha \cdot{\bm q}){\bm q}}{|{\bm q}|^4}
\label{berry_dip}
\end{equation}
where $\kappa_\pm=-1$ and $\kappa_0=2$ and the Berry dipole vector is defined as
\begin{equation}
{\bm d}_\alpha=-\xi_z(\xi_y s_\alpha,\xi_x c_\alpha,0),
\label{dip}
\end{equation}
with its direction that depends explicitly on the parameter $\alpha$.
Figure \ref{figure_dip} shows the Berry curvature and the corresponding Berry dipole of the conduction band near the four Dirac points ${\bm D}_{\xi_x=\pm 1,\xi_y=\pm 1,\xi_z=1}$. 

\section{Scattering through a Berry curvature domain wall}

In this section, we investigate the scattering properties of a 3D Dirac particle through a Berry curvature domain wall defined by a spatially dependent parameter  $\alpha(z)$, with $\alpha(z)=\alpha_L$ for $z<0$ and $\alpha(z)=\alpha_R$ for $z>0$.
This domain wall represents an abrupt change of the orientation of the Berry dipole $\bm d_\alpha$ on either side.
Our analysis proceeds in two complementary steps.
First, we employ a continuum description based on the effective low-energy Dirac Hamiltonian to elucidate the fundamental scattering mechanisms and the influence of Berry curvature discontinuities at the interface.
Second, we consider the lattice model to capture the discrete nature of the system and to incorporate additional scattering channels beyond those accessible within the continuum approximation.

From now on, to simplify the notation and whenever there is no ambiguity, we replace $\alpha_L$ and $\alpha_R$ by  $L$ and $R$, respectively.

\subsection{Low-energy continuum model description}
\label{sectionIIIA}
In the presence of the domain wall, the effective Hamiltonian rewrites
\begin{equation}
 H(z)=
\left\lbrace\begin{array}{ll}
H_{L}(z) & z<0,\\
H_{R}(z) & z>0.
     \end{array}\right.
     \label{simple}
\end{equation}
where
\begin{equation}
H_\alpha(z) =\hbar v_F(\hat{q}_x \Lambda_x+ \hat{q}_y \Lambda_y-i\xi_z\partial_z \Lambda_z^{\alpha}),
\end{equation}
with $\alpha=L,R$. This Hamiltonian describes the low-energy excitations around the Dirac point ${\bm D}_{\xi_x,\xi_y,\xi_z}$ [Eq. (\ref{dirac_point})].

Consider an incident wave of momentum ${\bm q}=(\hat{q}_x,\hat{q}_y,\hat{q}_z)$ and energy $E=\hbar v_F|{\bm q}|$. Since the transmitted momentum equals the incident momentum ${\bm q}$, and using translation invariance along the $xy-$plane, the scattering state can be written as $\Psi(x,y,z) =e^{i(\hat{q}_x x +\hat{q}_y y)}\Psi_{\theta,\varphi}(z)$ where the non trivial $z-$dependent part in the two regions takes the generic form
\begin{equation}
\begin{array}{ll}
\begin{split}
\Psi_{\theta,\varphi}(z)&=
\frac{1}{\sqrt{2}} \begin{pmatrix}
-\xi_x\sin\theta \cos\varphi-i\xi_zc_{L}\cos\theta \\
1\\
-\xi_y\sin\theta \sin\varphi+i\xi_zs_{L}\cos\theta  \end{pmatrix} e^{iq_z z}
\\&+\frac{r}{\sqrt{2}} \begin{pmatrix}
 -\xi_x\sin\theta \cos\varphi+i\xi_zc_{L}\cos\theta\\
1\\
-\xi_y\sin\theta \sin\varphi-i\xi_z s_{L}\cos\theta \end{pmatrix} e^{-iq_z z}, &z<0,
\end{split}
 \\
\Psi_{\theta,\varphi}(z)= \frac{t}{\sqrt{2}} \begin{pmatrix}
-\xi_x\sin\theta \cos\varphi-i\xi_z c_{R}\cos\theta\\
1\\
-\xi_y\sin\theta \sin\varphi+i\xi_z s_{R}\cos\theta\end{pmatrix} e^{iq_z z},  \hspace{0.6cm} z>0,
\end{array}
\label{wave3_junction}
\end{equation}
where $(q_x,q_y,q_z)=|{\bm q}|(\sin \theta\cos\varphi,\sin \theta\sin \varphi,\cos \theta)$.

To ensure conservation of the probability current perpendicular to the interface, the wavefunction amplitudes at the domain wall must satisfy an effective matching condition of the form
\begin{equation}
 \Lambda_z^{R}\Psi_{\theta,\varphi}(0^+)={\cal M}_{R L}\Lambda_z^{L}\Psi_{\theta,\varphi}(0^-),
 \label{matching}
\end{equation}
with ${\cal M}_{R L}$ a $3\times3$ matching matrix that also verifies
the following matrix equation:
\begin{equation}
{\cal M}_{R L}\Lambda_z^{L}{\cal M}_{R L}^{\dagger}=\Lambda_z^{R}.
\label{matchsxz}
\end{equation}
In order to determine the matching matrix $M_{RL}$,
a convenient approach is to {\it rotate} each Hamiltonian $H_\alpha$
 to the basis that diagonalizes $\Lambda_\alpha ^z$.
The corresponding orthogonal transformation is given by
\begin{equation}
O_{\alpha}=\frac{1}{\sqrt{2}}\begin{pmatrix}
c_{\alpha} & \sqrt{2}s_{\alpha}  & c_{\alpha}\\
i & 0 & -i\\
-s_{\alpha} & \sqrt{2}c_{\alpha}  & -s_{\alpha}
\end{pmatrix},
\end{equation}
such that the rotated Hamiltonian reads 
\begin{equation}
 {\bar H}_\alpha=O_{\alpha}^{\dag}  H_\alpha O_{\alpha}=\hbar v_F(-i\xi_z \partial_z S_z +\hat{q}_x{\bar \Lambda_x}^\alpha+\hat{q}_y{\bar \Lambda_y}^\alpha)
\end{equation}
with $S_z=O_{\alpha}^{\dag}  \Lambda_z^\alpha O_{\alpha}$ and ${\bar \Lambda_{x,y}}^\alpha=O_{\alpha}^{\dag}  \Lambda_{x,y} O_{\alpha}$. 
The scattering states in the rotated bases are given by $ {\bar \Psi}_{\theta,\varphi}(z)=O_{\alpha}^{\dag} \Psi_{\theta,\varphi}(z)$.
In this rotated basis, the matching condition [Eq. (\ref{matching})] transforms to
\begin{equation}
 S_z{\bar \Psi}_{\theta,\varphi}(0^+)={\bar{\cal M}}_{R L}S_z {\bar \Psi}_{\theta,\varphi}(0^-),
 \label{matchingz}
\end{equation}
where ${\bar{\cal M}}_{R L}= O_{R}^{\dag} {\cal M}_{R L} O_{L}$. 
Using a reduced singular value decomposition (SVD) \cite{Strang2009,Dwivedi2016}, the matrix $S_z$ can be decomposed as 
\begin{equation}
S_z=V \mathbb{1}_2 W^{\dag},
\label{svd_sz}  
\end{equation}
where 
\begin{equation}
V=\begin{pmatrix}
0& 1\\
0 & 0\\
-1 & 0
\end{pmatrix},\hspace{1cm} W=\begin{pmatrix}
0& 1\\
0 & 0\\
1 & 0
\end{pmatrix}, \hspace{1cm} \mathbb{1}_2=\begin{pmatrix}
1& 0\\
0 & 1
\end{pmatrix}
\label{vw}
\end{equation}
With this decomposition, the matching condition [Eq. (\ref{matchingz})] becomes
\begin{equation}
 \Phi_{\theta,\varphi}(0^+)={\cal K}_{R L}\Phi_{\theta,\varphi}(0^-),
 \label{matchingz2}
\end{equation}
where the reduced $2\times2$ matching matrix is defined by ${\cal K}_{R L}=V^{\dag}{\bar{\cal M}}_{R L}V $, and the transformed wavefunction is
 $\Phi_{\theta,\varphi}(z)=W^{\dag} {\bar \Psi}_{\theta,\varphi}(z)$.
To ensure conservation of the probability current, the matching matrix ${\cal K}_{R L}$ must satisfy the condition
\begin{equation}
{\cal K}_{R L}^{\dag} \sigma_z {\cal K}_{R L}=\sigma_z,
\label{cond match2}
\end{equation}
where $\sigma_z$ being the $z$ component of Pauli matrices.  
Following the approach of Ref.~\cite{Romeo2018}, we impose that the matching matrix ${\cal K}_{R L}$ belongs the special linear group $SL(2,C)$ which consists of $2\times2$ complex matrices with unit determinant. Under this requirement, the most general form of ${\cal K}_{R L}$ can be expressed as
\begin{equation}
\mathcal{K}_{R L}= \begin{pmatrix}
 e^{-i\lambda_1}\cosh x  &  -e^{-i\lambda_2}\sinh x \\
-e^{i\lambda_2}\sinh x   &  e^{i\lambda_1}\cosh x 
\end{pmatrix},
\label{genematch matr}
\end{equation}
with $x$ and $\lambda_{i}$ arbitrary real valued parameters.

As discussed in Ref.~\cite{Mandhour2025}, the general form of the matching matrix $\mathcal{K}_{R L}$ can be derived by introducing an explicit interface potential into the domain wall model. 
This effective low-energy scattering matrix $\mathcal{K}_{R L}$ implicitly assumes that the interface potential is sufficiently smooth so as to suppress intervalley scattering. An interesting open issue is to examine the effect of lattice mismatch at the interface, which would lead to effective Brillouin-zone folding and thereby induce intervalley scattering even at low energies, as discussed in Ref. \cite{Rodrigues2016}.

In the present work, we focus on the simplest case without any interface potential. Under these conditions, the matching matrix $\mathcal{K}_{R L}$ reduces to the identity matrix.
Applying the boundary condition given by Eq. (\ref{matchingz2}) with $\mathcal{K}_{R L}=\mathbb{1}_2$ 
we obtain the scattering amplitudes
\begin{subequations}
\begin{equation}
r=\frac{i \tan(\frac{\alpha_L-\alpha_R}{2}) \Delta_{R L}^{\xi_x\xi_y}(\theta,\varphi)}{2\xi_x\xi_y\xi_z\cos \theta - i \tan(\frac{\alpha_L-\alpha_R}{2}) \Delta_{R L}^{\xi_x\xi_y}(\theta,\varphi)},
\label{amplir_junction}
\end{equation}
and
\begin{equation}
t=\frac{2\xi_x\xi_y\xi_z\cos \theta}{2\xi_x\xi_y\xi_z\cos \theta - i \tan(\frac{\alpha_L-\alpha_R}{2}) \Delta_{R L}^{\xi_x\xi_y}(\theta,\varphi)}.
\label{amplit_junction}
\end{equation}
\end{subequations}
The transmission probability through the domain wall is then 
\begin{equation}
T_{\xi_x\xi_y}(\theta,\varphi)=|t|^2=\frac{4\cos^2\theta}{4\cos^2\theta+ \tan^2(\frac{\alpha_L-\alpha_R}{2}) \Delta_{R L}^{\xi_x\xi_y}(\theta,\varphi)^2},
\label{transstep}
\end{equation}
where 
\begin{equation}
\Delta_{R L}^{\xi_x\xi_y}(\theta,\varphi)=\frac{({\bm d}_R+{\bm d}_L) \cdot{\bm q}}{|\bm q|}\propto |{\bm \Omega}_{+}^R({\bm q})+{\bm \Omega}_{+}^L({\bm q})|,
 \label{berryjump}
\end{equation}
is the Berry dipole mismatch at the interface. Here ${\bm \Omega}_{+}^\alpha({\bm q})$ and $\bm d_\alpha$ ($\alpha=R,L$) are defined in Eqs.~(\ref{berry_dip}) and (\ref{dip}), respectively .
\begin{figure}[]
        \centering
    \includegraphics[width=0.5\textwidth]{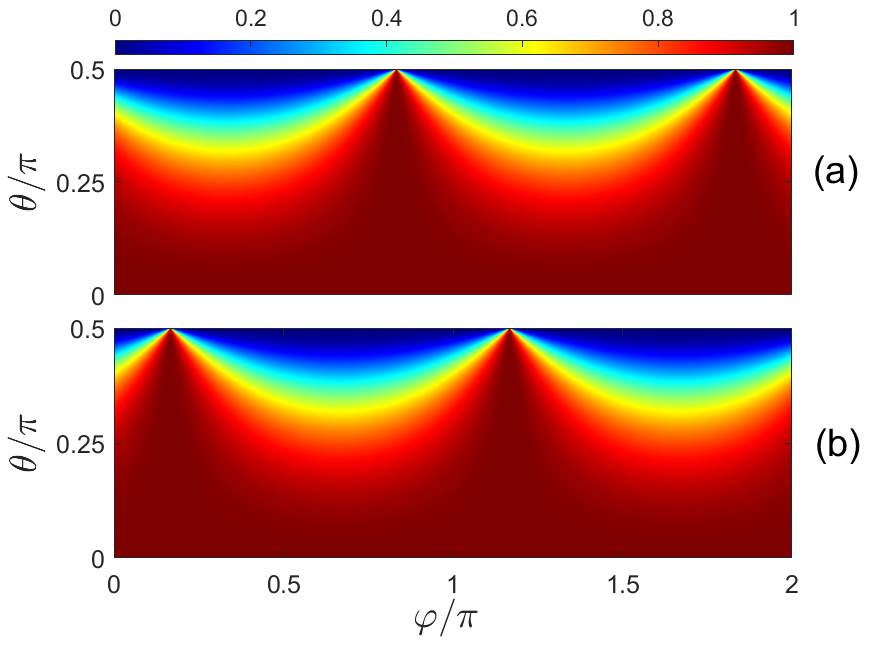} 
    \caption{Transmission probability $T_{\xi_x\xi_y}(\theta,\varphi)$ through the Berry dipole domain wall. $\alpha_L=0$ and $\alpha_R=\pi/3$. (a) $\xi_x\xi_y=1$ and (b) $\xi_x\xi_y=-1$}
\label{fig_cont}
\end{figure}
Unlike conventional scattering by a potential barrier, the transmission probability $T_{\xi_x\xi_y}(\theta,\varphi)$ is independent of the incident particle’s energy $E$. It depends solely on the incident angles $\theta$ and $\varphi$. Furthermore, the transmission exhibits valley dependency, resulting in two distinct expressions corresponding to $\xi_x\xi_y=\pm 1$. These two transmission probabilities are related by
\begin{equation}
T_{1}(\theta,\varphi)=T_{-1}(\theta,\varphi+\alpha_R+\alpha_L).
\end{equation}
This valley selectivity arises from the valley-dependent orientation of the Berry-curvature dipole vector, as illustrated in Fig. \ref{figure_dip}.
This splitting can be exploited for valley-selective transport or filtering.
From Eq. (\ref{transstep}), perfect transmission occurs when the Berry curvature flips its sign across the interface
\begin{equation}
\bm \Omega_+(\alpha_R,\theta,\varphi)=-\bm \Omega_+(\alpha_L,\theta,\varphi)
\label{perfect trans cond}
\end{equation}
which physically corresponds to the momentum ${\bm q}$ being orthogonal to the sum of the Berry dipole vectors on each side of the domain wall, ${\bm d}_R+{\bm d}_L$.
This condition is realized in three distinct cases.
The first corresponds to normal incidence ($\theta=0$) (see Fig. \ref{fig_cont}), which gives rise to the well-known Klein tunneling effect, as is well known for a potential barrier \cite{Allain2011,Illes2017}. The latter originates from the pseudospin conservation. For this incidence, the Berry curvature appears to vanish on both sides of the interface.
For oblique incidence, $\theta \neq 0$, two possible orientations of the in-plane wave vector ${\bm q_\parallel=(q_x,q_y)}$ satisfy the orthogonality condition with ${\bm d}_R+{\bm d}_L$. Such configurations produce perfect transmission for all $\theta \neq 0$ at two specific azimuthal angles $\varphi=-\xi_x\xi_y\frac{\alpha_R+\alpha_L}{2}+n \pi$, where $n$ is an integer (see Fig. \ref{fig_cont}).
Conversely, when the momentum ${\bm q}$ and the berry dipole vector $({\bm d}_R+{\bm d}_L)$ are collinear, 
the Berry curvature is conserved across the interface, $\bm \Omega_+(\alpha_R,\theta,\varphi)=\bm \Omega_+(\alpha_L,\theta,\varphi)$, and the transmission probability is minimized at the azimuthal angle $\varphi$ midway between those corresponding to perfect transmission, as depicted in Fig.~\ref{fig_cont}.
Physically, these results illustrate how the quantum geometric properties of Dirac particles, encoded in the Berry curvature and its dipole moment, govern the scattering properties of the domain wall.

\subsection{Lattice description}
\label{sectionIIIB}

Hereafter we consider inhomogeneous Hamiltonian models that describes an interface,
parallel to $xy$-plane and located at position $n_z=0$, separating two domains characterized by distinct parameters $\alpha_L$  (left of the domain wall)  for $n_z< 0$ and $\alpha_R$  (right of the domain wall)  for $n_z\geq0$.
The effective lattice Hamiltonian for this domain wall is given by
 $H=H_{L}(n_z< 0)+H_{R}(n_z\geq 0)$.

\subsubsection{Effective one dimensional model}

\begin{figure}[h!]
\setlength{\unitlength}{1mm}
\includegraphics[width=0.5\textwidth]{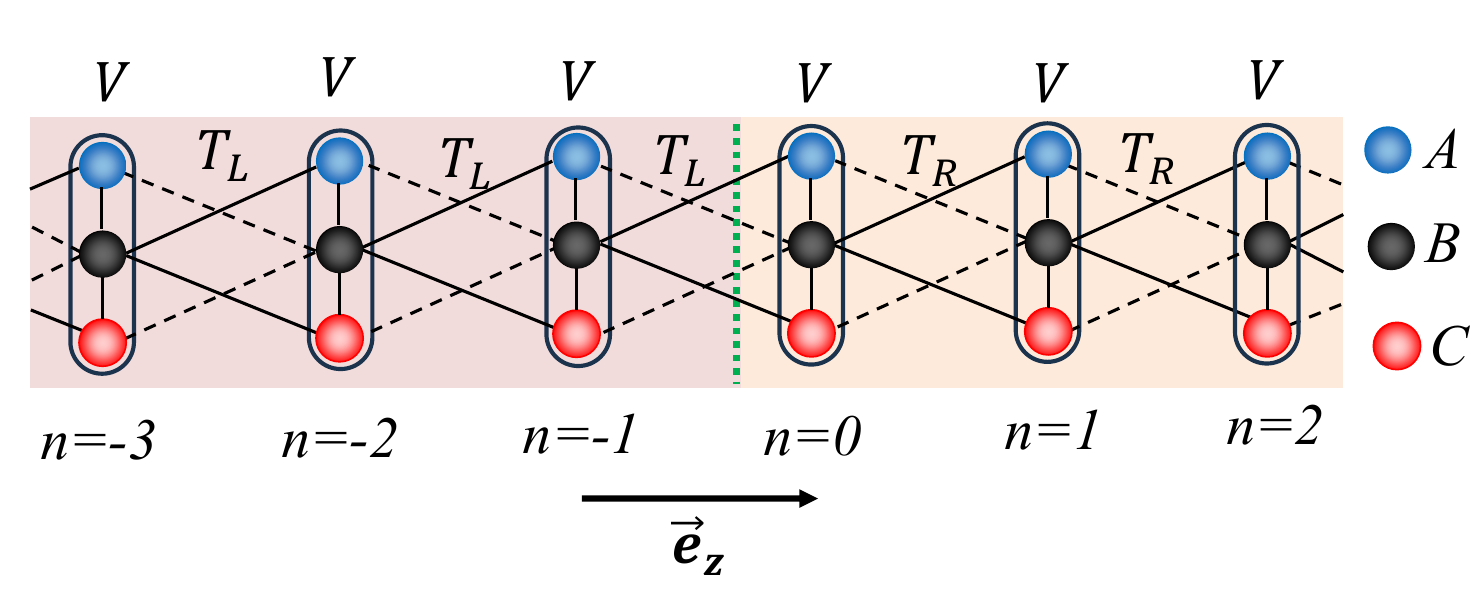}
\caption{Effective one-dimensional model picture of the domain walls.}
\label{fig1d}
\end{figure}

For a domain wall perpendicular to the $z$ axis,the Hamiltonian remains translation invariant along the directions parallel to the interface $(\bm e_x,\bm e_y)$. We may therefore apply Bloch’s theorem in the $x$ and $y$ directions and define a two-dimensional Bloch basis of creation operators
$X^{\dagger}_{n_z}(k_x,k_y)=
 \sum_{n_x,n_y} e^{i (n_x k_x +n_y k_y)}X^{\dagger}_{n_x,n_y,n_z}$ with $X=A,B,C$.
Introducing the three-component creation operators $d_{n}^{\dagger}(k_x,k_y)\equiv (A^{\dagger}_{n},B^{\dagger}_{n},C^{\dagger}_{n})$ ($n\equiv n_z$)
the  homogeneous model $H_\alpha$ can be written as $H_\alpha=\int_{BZ} \dfrac{dk_xdk_y}{(2\pi)^2} H_\alpha(k_x,k_y)$ where $H_\alpha(k_x,k_y)$ defines a $(k_x,k_y)$-dependent effective one dimensional tight-binding model, taking the form
\begin{equation}
H_\alpha(k_x,k_y)=\sum_n d_{n}^{\dagger}V d_{n}+d_{n}^{\dagger }T_\alpha^{\dagger} d_{n-1} + d_{n-1}^{\dagger}T_\alpha d_{n},
\label{one chain}
\end{equation}
with onsite and nearest-neighbor hopping matrices given respectively by
\begin{equation}
V(k_x,k_y)= t\left(
 \begin{array}{ccc}
  0& c_x &0\\
  c_x&0&c_y\\
  0&c_y&0
 \end{array}\right), \hspace{0.1 cm} T_\alpha=\frac{t}{2}\left(
 \begin{array}{ccc}
  0&  -c_\alpha &0\\
   c_\alpha &0&  -s_\alpha\\
  0& s_\alpha &0
 \end{array}\right).
  \label{valpha}
 \end{equation}
In the presence of a domain wall at $n=n_z=0$ the effective one dimensional Hamiltonian then  writes (see Fig. \ref{fig1d})
\begin{equation}
 H(k_x,k_y)=H_{L}(k_x,k_y, n< 0)+H_{R}(k_x,k_y, n\geq 0).
\end{equation}

\subsubsection{Scattering properties of the domain wall perpendicular to z axis}

We now analyze the scattering properties of the lattice model for a domain wall
perpendicular to z axis.
Specifically, we search for scattering state $\Psi^{\dagger}(\varepsilon,k_x,k_y)$ of energy $E=\varepsilon t$ and transverse momentum $k_x,k_y$, which can be expanded as
$\Psi^{\dagger}=\sum_n \sum_{X}\psi_{n}^X(\varepsilon,k_x,k_y) X^{\dagger}_{n}(k_x,k_y)$
($X=A,B,C$).
These states satisfy the eigenvalue equation $H(k_x,k_y)\Psi^{\dagger}=t\varepsilon \Psi^{\dagger}$.
In terms of the amplitude vectors at each lattice site, defined as $|\Psi_{n}\rangle \equiv \left(\psi_{n}^A, \psi_{n}^B, \psi_{n}^C\right)^T$, 
this eigenvalue constraint translates into a set of coupled equations 
$|\Psi_{n}\rangle \equiv \left(\psi_{n}^A, \psi_{n}^B, \psi_{n}^C\right)^T$
\begin{equation}
 \begin{array}{ll}
\begin{array}{l}
 n<0, \alpha=L\\
n> 0, \alpha=R\\
\end{array}
  &M|\Psi_{n}\rangle =T_\alpha |\Psi_{n+1}\rangle+T_\alpha^\dag |\Psi_{n-1}\rangle,\\
& \\
n=0 &M|\Psi_{0}\rangle =T_R |\Psi_{1}\rangle+T_L^\dag |\Psi_{-1}\rangle,\\
\end{array}
 \label{eq scat}
\end{equation}
where $M(\varepsilon,k_x,k_y)=\varepsilon-V(k_x,k_y)$.

The scattering problem is conveniently addressed in the basis that diagonalizes the hopping matrix $T_\alpha$.
The corresponding orthogonal transformation is given by
\begin{equation}
O_{\alpha}=\frac{1}{\sqrt{2}}\begin{pmatrix}
c_{\alpha} & \sqrt{2}s_{\alpha}  & c_{\alpha}\\
-i & 0 & i\\
-s_{\alpha} & \sqrt{2}c_{\alpha}  & -s_{\alpha}
\end{pmatrix},
\end{equation}
such that $2O_{\alpha}^{\dag}  T_\alpha O_{\alpha}=iS_z$, where $S_z$ denotes the  $z-$components of the pseudospin-1 operator. In this transformed basis, Eq. (\ref{eq scat}) takes the form
\begin{equation}
 \begin{array}{ll}
\begin{array}{l}
 n<0, \alpha=L\\
n>0, \alpha=R\\
\end{array}
  &\hat{M}_\alpha|\Phi_{n}\rangle=iS_z(|\Phi_{n+1}\rangle-|\Phi_{n-1}\rangle).\\
& \\
n=0  &\hat{M}_{RL}|\Phi_{0}\rangle=iS_z|\Phi_{1}\rangle-O_{R}^{\dag}O_{L}iS_z|\Phi_{-1}\rangle,\\
\end{array}
 \label{eq2z}
\end{equation}
where the operators are defined as
\begin{equation}
\hat{M}_\alpha=2O_{\alpha}^{\dag}  M O_{\alpha}, \hspace{0.2cm} \hat{M}_{RL}=2O_{R}^{\dag}  M O_{L},
\end{equation}
and the transformed wavefunction amplitudes read
\begin{equation}
\begin{array}{l}
 n<0, \alpha=L\\
n \geq 0, \alpha=R\\
\end{array}
\hspace{0.5cm}|\Phi_{n}\rangle=O_{\alpha}^{\dag}|\Psi_{n}\rangle \equiv \left(\alpha_{n},\gamma_n , \beta_n \right)^T.
 \end{equation}

Using the expression of $S_z$ given by Eq. (\ref{svd_sz}), the recursion relations in Eq.~(\ref{eq2z}) can be rewritten as

\begin{equation}
 \begin{array}{ll}
\begin{array}{l}
 n<0, \alpha=L\\
n>0, \alpha=R\\
\end{array}
  &\mathcal{G}_\alpha^{-1}|A_n\rangle+|A_{n-1}\rangle=|A_{n+1}\rangle.\\
& \\
n=0  &\mathcal{G}_R^{-1}|A_0\rangle+\Lambda_{RL}|A_{-1}\rangle=|A_{1}\rangle.\\
\end{array}
\label{eqred}
\end{equation}
Here, $\mathcal{G}_\alpha=iW^\dag \hat{M}_\alpha^{-1}V$ and $\Lambda_{RL}=i\mathcal{G}^{-1}_\alpha W^\dag  \hat{M}_R^{-1} O_{R}^{\dag}O_{L} V$ are $2\times 2$ matrices, where $V$ and $W$ are defined in Eq. (\ref{vw}). Explicitly, 
\begin{equation}
\mathcal{G}^{-1}_\alpha=\frac{2i}{\varepsilon} \begin{pmatrix}
\varepsilon^2-u_\alpha^2 & u_\alpha^2-i\varepsilon v_\alpha\\
-u_\alpha^2-i\varepsilon v_\alpha & u_\alpha^2-\varepsilon^2
\end{pmatrix}
\end{equation}
and 
\begin{equation}
\Lambda_{RL}=\frac{1}{2}\begin{pmatrix}
\cos\Delta \alpha+1+iX &-\cos\Delta \alpha+1+iX\\
-\cos\Delta \alpha+1-iX & \cos\Delta \alpha+1-iX
\end{pmatrix}
\end{equation}
where $X=\frac{\sqrt{2}u_R}{\varepsilon} \sin \Delta \alpha$ with $\Delta \alpha=\alpha_R-\alpha_L$.
The momentum-dependent functions $u_\alpha$ and $v_\alpha$ are defined as
\begin{equation}
\begin{array}{cc}
u_\alpha(k_x,k_y)=&\frac{1}{\sqrt{2}}(s_\alpha c_x+c_\alpha c_y), \\
v_\alpha(k_x,k_y)=&c_\alpha c_x-s_\alpha c_y.
\end{array}
\label{uv}
\end{equation}
The two-component wave function $|A_n\rangle$ is obtained by projecting the transformed state as
\begin{equation}
|A_n\rangle=W^\dag |\Phi_{n}\rangle \equiv \left(\beta_{n}, \alpha_n \right)^T,
\end{equation}
Introducing the four-component vector
\begin{equation}
|\Omega_{n}\rangle\equiv \left(\beta_{n}, \alpha_n,  \beta_{n-1}, \alpha_{n-1} \right)^T,
\label{state}
\end{equation}
the recursion relations (\ref{eqred}) can be reformulated as
\begin{equation}
 \begin{array}{ll}
\begin{array}{l}
 n<0, \alpha=L\\
n>0, \alpha=R\\
\end{array}
  &\mathbb{T}_\alpha|\Omega_{n}\rangle=|\Omega_{n+1}\rangle.\\
& \\
n=0  &\mathbb{T}_{RL}|\Omega_{0}\rangle=|\Omega_{1}\rangle.\\
\end{array}
\label{t transfert}
\end{equation}
where the transfer matrices are
\begin{equation}
\mathbb{T}_\alpha=\begin{pmatrix}
\mathcal{G}^{-1}_\alpha & \mathbb{1}_2\\
\mathbb{1}_2 & 0_2
\end{pmatrix},
 \hspace{1cm} 
 \mathbb{T}_{RL}=\begin{pmatrix}
\mathcal{G}^{-1}_R & \Lambda_{RL}\\
\mathbb{1}_2 & 0_2
\end{pmatrix},
\end{equation}
with $\mathbb{1}_2$ and $0_2$ the $2 \times 2$  identity and zero matrices, respectively.

Remarkably, the matrices $\mathbb{T}_\alpha$ with $\alpha= L,R$ have eigenvalues $\lambda_{ss'}$ ($s,s'=\pm$) that do not depend on $\alpha$:
\begin{equation}
\lambda_{ss'}=se^{is'k_z}.
\end{equation}
Here, the energy is chosen in the band spectrum $\varepsilon=s_z^2+2u_\alpha^2+v_\alpha^2$ with $s_z=\sin k_z$ and $u_\alpha,v_\alpha$ defined in Eq. (\ref{uv}).
In contrast to the eigenvalues $\lambda_{ss'}$, the corresponding eigenvectors explicitly depend on $\alpha$ and are given by
\begin{equation}
|\alpha, s,s'\rangle \equiv \left(X_+,X_-,\lambda_{ss'}^*X_+,\lambda_{ss'}^*X_- \right)^T
\end{equation}
where $X_\pm=v_\alpha+i(ss's_z\pm\varepsilon)$.
For $s'=+$, ($s'=-$) there are two right-moving (left-moving) plane-wave eigenvectors, labeled $s=\pm$.

\begin{figure}[]
\setlength{\unitlength}{1mm}
\includegraphics[width=0.5\textwidth]{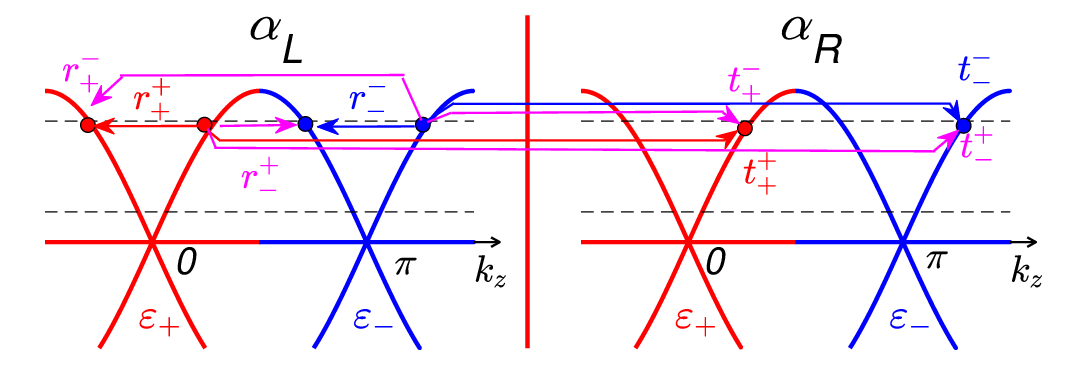}
\includegraphics[width=0.5\textwidth]{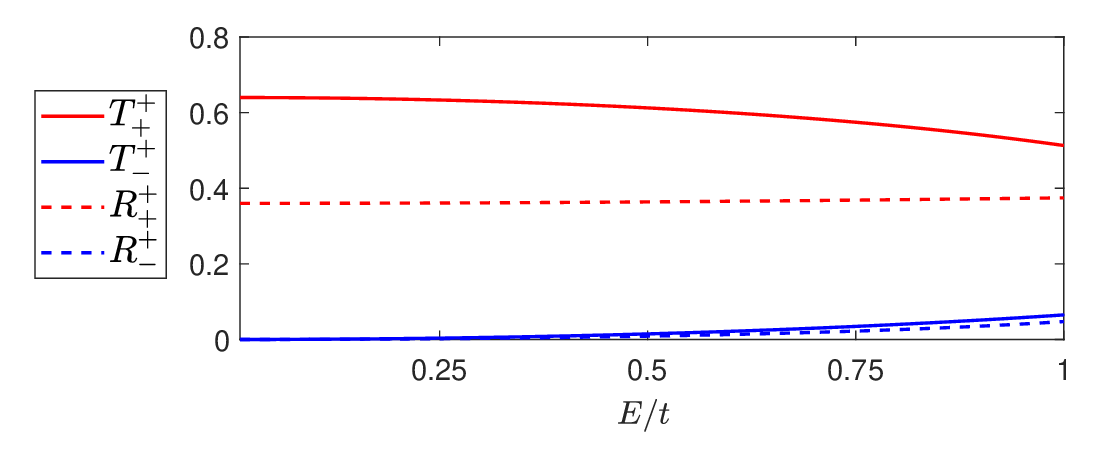}
\caption{Top panel: Schematic of transmission and reflection probabilities across the Berry dipole domain wall. Bottom panel: intervalley $(T_-^+,R_-^+)$ and intravalley $(T_+^+,R_+^+)$ transmission and reflection probabilities computed for $\alpha_L=0$, $\alpha_R=\pi/3$, $\theta=\pi/3$ and $\varphi=\pi/6$.}.
\label{fig_trans}
\end{figure}

For a given energy $\varepsilon>0$ and momentums $k_x$, $k_y$, the wave function given by Eq. (\ref{state}) can be expressed as
\begin{equation}
 |\Omega_n\rangle=\left \lbrace
 \begin{array}{lr}
  \sum_{s,s'=\pm} a_{ss'} \lambda_{ss'}^n |L, s,s'\rangle  & n< 0,\\
  \sum_{s,s'=\pm} b_{ss'} \lambda_{ss'}^n |R, s,s'\rangle  & n\geq 0.
 \end{array}\right.
\end{equation}
Since there are two incident right-moving waves, $|L, s=\pm, s'=+\rangle$, the boundary condition $|\Omega_{1}\rangle= \mathbb{T}_{RL}| \Omega_{0}\rangle$ can be written for each right-moving mode as
\begin{equation}
\begin{array}{ll}
\mathbb{T}_{RL}&[|L, s,+\rangle+r_{+}^s|L, +,-\rangle+r_{-}^s|L, -,-\rangle]\\
&=t_{+}^s\lambda_{++} |R, +,+\rangle+t_{-}^s\lambda_{-+} |R, -,+\rangle,
\end{array}
\end{equation}
where $s=\pm$ labels the two incident modes.
For an incident right-moving wave $|L, s,+\rangle$ in region $L$, two reflected left-moving waves $|L, \pm,-\rangle$ arise with corresponding amplitudes $r_\pm^s$, along with two transmitted right-moving waves in region $R$ with amplitudes $t_\pm^s$.
Our numerical results show that the total transmission probability is identical for each incident mode, and is given by
\begin{equation}
T_{\textrm{latt}}(k_x,k_y,\varepsilon)=|t_{+}^+|^2+|t_{-}^+|^2=|t_{+}^-|^2+|t_{-}^-|^2.
\label{trans_mode}
\end{equation}
Moreover, in the low-energy limit, intermode scattering vanishes, i.e., $t_+^- = t_-^+ = r_+^- = r_-^+ = 0$
so that each incident mode is transmitted independently without mixing between modes.
These two modes originate from the doubling of the unit cell. As shown in Eq. (\ref{state}), the unit cell contains two sites, resulting in a doubling along the $z$-direction. 
Consequently, the first Brillouin zone is halved, and the energy bands are doubled \cite{Rodrigues2012,Rodrigues2013}.
 We denote  $\bm {e}_x$, $\bm {e}_y$ and $2\bm {e}_z$ the Bravais lattice vectors, instead of $\bm {e}_x$, $\bm {e}_y$ and $\bm {e}_z$ of the pristine lattice. The corresponding reciprocal lattice vectors are then $2\pi\bm {e}_x$, $2\pi\bm {e}_y$ and $\pi\bm {e}_z$. 
The unfolded first Brillouin zone is a cube with $-\pi \leq k_x, k_y, k_z \leq \pi$, while the folded Brillouin zone is a rectangular prism with $-\pi \leq k_x, k_y\leq \pi$ and $-\pi/2 \leq  k_z \leq \pi/2$.
The two bands in the folded Brillouin zone are given by
\begin{equation}
\varepsilon_+=\varepsilon(\bm{k}), \hspace{1cm} \varepsilon_-=\varepsilon(\bm{k}+\pi\bm{e}_z),
\end{equation}
which are identical.
For a given transverse wave vectors $k_x$ and $k_y$ and an energy $\varepsilon>0$ in the spectrum, there exist four possible longitudinal wave vectors $k_z$ given by
\begin{equation}
k_z=\pm q_o, \hspace{1cm} k_z=\pm \left(q_o+\pi \right),
\label{modes}
\end{equation}
where $q_o=\arcsin \sqrt{\varepsilon^2-\cos^2k_x-\cos^2k_y}$. 
These four wave vectors are real, resulting in two distinct propagation modes. The first mode, with $k_z = q_0$, originates from the $\varepsilon_+$ band, whereas the second mode, with $k_z = q_0 + \pi$, originates from the $\varepsilon_-$ band, as illustrated in Fig. \ref{fig_trans}. Because these bands are identical, the total transmission probability is the same for each incident mode, as given in Eq. (\ref{trans_mode}). 
Physically, the two bands $\varepsilon_\pm$ correspond to a pair of Dirac cones $\bm D_{\xi_x,\xi_y,\pm}$ with identical in-plane momentum $(k_x,k_y)$ but separated along the $k_z$ direction. These distinct $k_z$ positions define the propagation modes given by Eq. (\ref{modes}).
As shown in Fig. \ref{fig_trans}, the large separation of these modes in reciprocal space at low energies strongly suppresses intervalley scattering. 
However, at higher energies, the modes approach each other, slightly enhancing the coupling between states in the two Dirac cones and leading to increased intervalley scattering.

\begin{figure}[]
        \centering
      \includegraphics[width=0.5\textwidth]{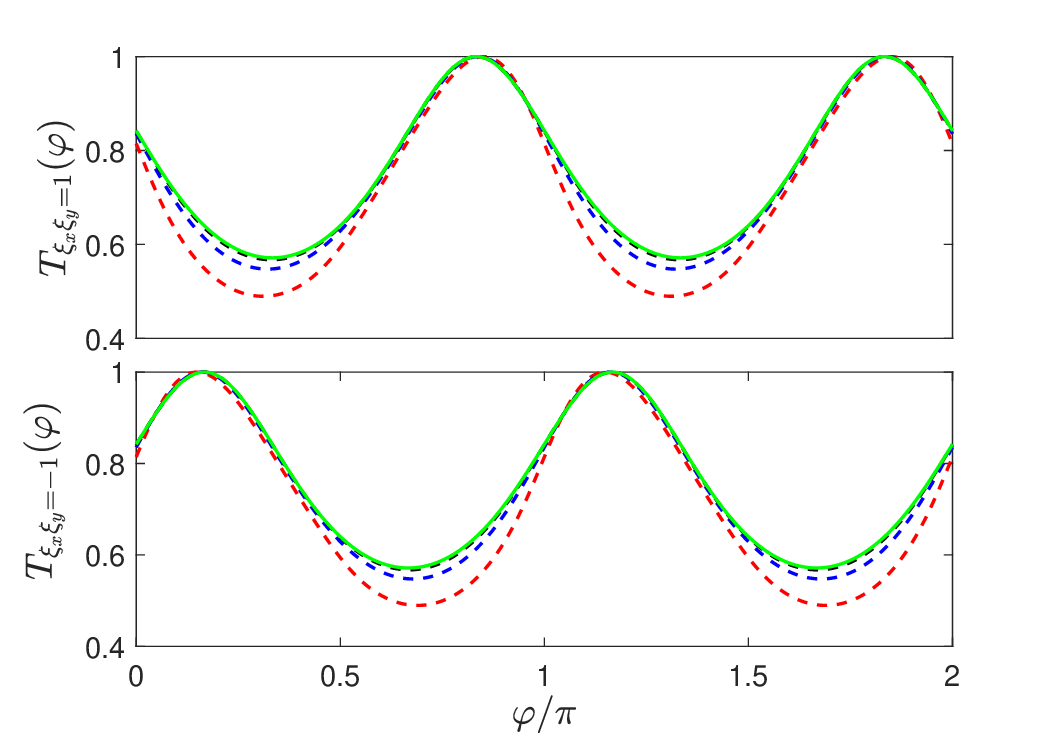}
\caption{Transmission probabilities through the domain wall in term of the incidence angle $\varphi$ for $\theta=\pi/3$ and for different values of the energy $\varepsilon$. The upper panel corresponds to the incident wave vector around the Dirac points $D_{1,1,0}$ and $D_{-1,-1,0}$, and the lower panel to $D_{1,-1,0}$ and $D_{-1,1,0}$.  
Green solid lines $T_{\textrm{cont}}$ for the low energy continuum description and dashed lines $T_{\textrm{latt}}$ for the lattice description. $\alpha_L=0$ and $\alpha_R=\pi/3$.
Dashed red line $\varepsilon=1.1$, dashed blue line $\varepsilon=0.7$ and dashed black line $\varepsilon=0.3$.}
\label{figure1}
\end{figure}

For a given energy $\varepsilon>0$ and a wave vector $\bm k(k_x,k_y,k_z)$ around the Dirac point $\bm D_{\xi_x,\xi_y,1}=(\xi_x\pi/2,\xi_y\pi/2,0)$, we can rewrite
$\bm k=\bm D_{\xi_x,\xi_y,1}+\bm q$, $\bm q=(-\xi_xq_x,-\xi_yq_y,q_z)$ with $q_x=q \sin \theta \cos \varphi$, and $q_y=q \sin \theta \sin \varphi$,
where the angles $0 \leq \theta \leq \pi/2$ and $0 \leq \varphi \leq 2\pi$ define the direction of incidence. More explicitly, the relations between the incidence angles $(\theta, \varphi)$ and the in-plane momenta $(q_x, q_y)$ are
\begin{equation}
\varphi=\arctan \frac{q_y}{q_x}, \hspace{1cm} \theta=\arctan \frac{\sqrt{q_x^2+q_y^2}}{q_z},
\label{ky teta1}
\end{equation}
where $q_z$ is determined by the energy constraint
\begin{equation}
q_z=\arcsin \sqrt{\varepsilon^2-\sin^2 q_x-\sin^2 q_y}
\end{equation}
Consequently, the transmission probability $T_{\textrm{latt}}(\varepsilon,\theta,\varphi)$ can be obtained parametrically as a function of the energy
$\varepsilon$ and the incidence angles $(\theta,\varphi)$. 
Figure~\ref{figure1} displays the comparison between $T_{\textrm{latt}}(\varepsilon,\theta,\varphi)$ and the continuum model transmission $T_{\textrm{cont}}(\theta,\varphi)$ for several values of the energy $\varepsilon$. 
At low energies, the lattice and continuum models agree quantitatively, confirming the validity of the continuum approximation.


\section{Conclusion}
  
In this work, we investigate the scattering of three-dimensional massless Dirac particles across a domain wall that separates two regions with distinct quantum geometries characterized by differing Berry-curvature dipole orientations. Using a tunable three-band multifold Hopf-semimetal model, we find that a mismatch in dipole orientation across the interface induces partial reflection and transmission, while both the transmitted momentum and particle energy remain conserved through the domain wall.

In a first step, we employ a low-energy continuum description and demonstrate that the transmission probability depends solely on the incident angles and the Berry-curvature vector mismatch at the domain-wall interface, while remaining independent of the particle’s energy. Notably, perfect transmission occurs at normal incidence regardless of the Berry-curvature mismatch. Conversely, the mismatch induces perfect transmission at specific oblique incidence angles. Additionally, the transmission exhibits valley selectivity as a result of the valley-dependent orientation of the Berry-curvature dipole vector.

In a second step, we examine the scattering properties within a tight-binding description on the cubic lattice. In this regime, the transmission probability depends on the Berry curvature mismatch at the interface, the incident angle, and the particle energy. Unlike the low-energy continuum limit, higher-energy 3D Dirac particles propagate through the domain wall via two distinct scattering modes. These modes originate from the doubling of the unit cell along the $z$-axis, which folds the Brillouin zone and results in a pair of energy bands participating in the scattering process.
At sufficiently low energies, the two modes become well separated, suppressing intermode scattering, and the transmission probability exhibits excellent quantitative agreement with the continuum model for all angles of incidence.

This geometric scattering model fundamentally differs from conventional scattering mechanisms, as it induces pseudospin scattering despite identical incident and refracted momenta, thereby highlighting the intrinsic role of quantum geometry independent of momentum transfer.

Our results demonstrate that spatial variations in quantum geometry constitute a distinct scattering mechanism for Dirac particles, fundamentally different from conventional impurity or potential barrier scattering. This insight suggests the possibility to tune transport properties in quantum materials via engineered quantum geometric textures.

\label{sectionIV}

\section*{Acknowledgments}
L.M. acknowledges the LPS in Orsay for financial support and kind hospitality, where the work started.

\end{document}